\documentstyle[11pt]{article}
\renewcommand{\section}[1]{\addtocounter{section}{1}
       \vspace{5mm} \par \noindent
        {\bf \thesection . #1}\setcounter{subsection}{0}
        \par \vspace{2mm} }
\renewcommand{\subsection}[1]{\addtocounter{subsection}{1}
         \vspace{2.5mm}\par\noindent {\em \thesubsection . #1}\par
          \vspace{0.5mm} }
\renewcommand{\thebibliography}[1]{
          { \vspace{5mm}\par\noindent
          {\bf References}\par\vspace{4mm} } \list{[\arabic{enumi}]}
          {\settowidth\labelwidth{[#1]}\leftmargin
          \labelwidth \advance\leftmargin\labelsep\addtolength{\topsep}{-4em}
          \usecounter{enumi}}  }

\def\eqref#1{(\ref{#1})}

\def\a{\alpha}
\def\b{\beta}
\def\g{\gamma}
\def\G{\Gamma}
\def\d{\delta}
\def\e{\epsilon}

\def\s{\sigma}
\def\S{\Sigma}
\def\t{\theta}

\def\1d{\dot 1}
\def\2d{\dot 2}
\def\nn{\nonumber}

\def\la{\label}
\def\bm{\bibitem}
\def\be{\begin{equation}}
\def\ee{\end{equation}}
\def\bea{\begin{eqnarray}}
\def\eea{\end{eqnarray}}
\def\ed{\end{document}}
\def\qq{\quad\quad}

\def\[{{[}}
\def\]{{]}}

\def\h{{1\over {\sqrt \lambda}}}
\def\o{{1\over \lambda}}

\begin{document}

{\baselineskip=14pt
\begin{flushright}
\hfill{UMDEPP 97-001}\\
\hfill{CTP TAMU-25/96}\\
\hfill{IC/96/124}\\
\hfill{hep-th/9607185}\\
\end{flushright}
}


\def\a{\alpha} \def\b{\beta} \def\c{\chi} \def\d{\delta}
\def\e{\epsilon} \def\f{\phi} \def\g{\gamma} \def\h{\eta}
\def\i{\iota} \def\j{\psi} \def\k{\kappa} \def\l{\lambda}
\def\m{\mu} \def\n{\nu} \def\o{\omega} \def\p{\pi}
\def\q{\theta} \def\r{\rho} \def\s{\sigma} \def\t{\tau}
\def\u{\upsilon} \def\x{\xi} \def\z{\zeta} \def\D{\Delta}
\def\F{\Phi} \def\G{\Gamma} \def\J{\Psi} \def\L{\Lambda}
\def\O{\Omega} \def\P{\Pi} \def\Q{\Theta} \def\S{\Sigma}
\def\U{\Upsilon} \def\X{\Xi}

\def\half{{\fracm12}}
\def\scst{\scriptstyle}
\def\dsl{{D\!\!\!\! /}}
\def\nsl{{n\!\!\! /}\,}
\def\-{{\hskip 1.5pt}\hbox{-}}

\def\doit#1#2{\ifcase#1\or#2\fi}

\def\frac#1#2{{\textstyle{#1\over\vphantom2\smash{\raise.20ex
        \hbox{$\scriptstyle{#2}$}}}}}                   
\def\fracm#1#2{\,\hbox{\large{${\frac{{#1}}{{#2}}}$}}\,}
\def\fracmm#1#2{\,{{#1}\over{#2}}\,}

\def\Bar#1{\overline{#1}}                       
\def\leftrightarrowfill{$\mathsurround=0pt \mathord\leftarrow \mkern-6mu
        \cleaders\hbox{$\mkern-2mu \mathord- \mkern-2mu$}\hfill
        \mkern-6mu \mathord\rightarrow$}

\newskip\humongous \humongous=0pt plus 1000pt minus 1000pt
\def\caja{\mathsurround=0pt}
\def\eqalign#1{\,\vcenter{\openup2\jot \caja
        \ialign{\strut \hfil$\displaystyle{##}$&$
        \displaystyle{{}##}$\hfil\crcr#1\crcr}}\,}
\newif\ifdtup
\def\panorama{\global\dtuptrue \openup2\jot \caja
        \everycr{\noalign{\ifdtup \global\dtupfalse
        \vskip-\lineskiplimit \vskip\normallineskiplimit
        \else \penalty\interdisplaylinepenalty \fi}}}
\def\li#1{\panorama \tabskip=\humongous                         
        \halign to\displaywidth{\hfil$\displaystyle{##}$
        \tabskip=0pt&$\displaystyle{{}##}$\hfil
        \tabskip=\humongous&\llap{$##$}\tabskip=0pt
        \crcr#1\crcr}}

\def\[{\lfloor{\hskip 0.35pt}\!\!\!\lceil\,}
\def\]{\,\rfloor{\hskip 0.35pt}\!\!\!\rceil}
\def\du#1#2{_{#1}{}^{#2}}
\def\ud#1#2{^{#1}{}_{#2}}
\def\dud#1#2#3{_{#1}{}^{#2}{}_{#3}}
\def\udu#1#2#3{^{#1}{}_{#2}{}^{#3}}

\def\pl#1#2#3{Phys.~Lett.~{\bf {#1}B} (19{#2}) #3}
\def\np#1#2#3{Nucl.~Phys.~{\bf B{#1}} (19{#2}) #3}
\def\prl#1#2#3{Phys.~Rev.~Lett.~{\bf #1} (19{#2}) #3}
\def\pr#1#2#3{Phys.~Rev.~{\bf D{#1}} (19{#2}) #3}
\def\cqg#1#2#3{Class.~and Quant.~Gr.~{\bf {#1}} (19{#2}) #3}
\def\cmp#1#2#3{Comm.~Math.~Phys.~{\bf {#1}} (19{#2}) #3}
\def\jmp#1#2#3{Jour.~Math.~Phys.~{\bf {#1}} (19{#2}) #3}
\def\ap#1#2#3{Ann.~of Phys.~{\bf {#1}} (19{#2}) #3}
\def\prep#1#2#3{Phys.~Rep.~{\bf {#1}C} (19{#2}) #3}
\def\ptp#1#2#3{Prog.~Theor.~Phys.~{\bf {#1}} (19{#2}) #3}
\def\ijmp#1#2#3{Int.~Jour.~Mod.~Phys.~{\bf A{#1}} (19{#2}) #3}
\def\nc#1#2#3{Nuovo Cim.~{\bf {#1}} (19{#2}) #3}
\def\ibid#1#2#3{{\it ibid.}~{\bf {#1}} (19{#2}) #3}
\def\grg#1#2#3{Gen.~Rel.~Grav.~{\bf{#1}} (19{#2}) {#3} }
\def\pla#1#2#3{Phys.~Lett.~{\bf A{#1}} (19{#2}) {#3}}
\def\mpl#1#2#3{Mod.~Phys.~Lett.~{\bf A{#1}} (19{#2}) #3}
\def\zp#1#2#3{Zeit.~f\"ur Phys.~{\bf{#1}C} (19{#2}) {#3}}
\def\jgtp#1#2#3{Jour.~of Group Theory for Physicists, {\bf{#1}} (19{#2}) {#3}}
\def\rmp#1#2#3{Rev.~Mod.~Phys.~{\bf {#1}} (19{#2}) {#3}}

\def\fracmm#1#2{{{#1}\over{#2}}}
\def\frac#1#2{{\textstyle{#1\over\vphantom2\smash{\raise -.20ex
        \hbox{$\scriptstyle{#2}$}}}}}                   
\def\fracm#1#2{\hbox{\large{${\frac{{#1}}{{#2}}}$}}}

\def\Dot#1{\buildrel{_{_{\hskip 0.01in}{\normalsize\bullet}}}\over{#1}}
\def\Tilde#1{{\widetilde{#1}}\hskip 0.015in}
\def\Hat#1{\widehat{#1}}
\def\low#1{{\raise -3pt\hbox{${\hskip 1.0pt}\!_{#1}$}}}

\def\tr{{\rm tr}}
\def\Tr{{\rm Tr}}
\def\ZZ{\rlx\leavevmode
             \ifmmode\mathchoice
                    {\hbox{\cmss Z\kern-.4em Z}}
                    {\hbox{\cmss Z\kern-.4em Z}}
                    {\lower.9pt\hbox{\cmsss Z\kern-.36em Z}}
                    {\lower1.2pt\hbox{\cmsss Z\kern-.36em Z}}
               \else{\cmss Z\kern-.4em Z}\fi}
\def\IC{\rlx\leavevmode
             \ifmmode\mathchoice
                    {\hbox{\kern.33em\inbar\kern-.3em{\rm C}}}
                    {\hbox{\kern.33em\inbar\kern-.3em{\rm C}}}
                    {\hbox{\kern.28em\sinbar\kern-.25em{\rm C}}}
                    {\hbox{\kern.25em\ssinbar\kern-.22em{\rm C}}}
             \else{\hbox{\kern.3em\inbar\kern-.3em{\rm C}}}\fi}

\newbox\leftpage \newdimen\fullhsize \newdimen\hstitle \newdimen\hsbody
\tolerance=1000\hfuzz=2pt\def\fontflag{cm}
\catcode`\@=11 
\hsbody=\hsize \hstitle=\hsize 
\def\almostshipout#1{\if L\l@r \count1=1 \message{[\the\count0.\the\count1]}
      \global\setbox\leftpage=#1 \global\let\l@r=R
 \else \count1=2
  \shipout\vbox{\speclscape{\hsize\fullhsize\makeheadline}
      \hbox to\fullhsize{\box\leftpage\hfil#1}}  \global\let\l@r=L\fi}

\def\items#1{\\ \item{[#1]}}
\def\eqntwo#1#2{eqs.\,(\ref{#1}) and (\ref{#2})}
\def\eqnthree#1#2#3{eqs.\,(\ref{#1}), (\ref{#2}) and (\ref{#3})}

\def\Dot#1{\buildrel{\hskip1.5pt\raise-0.03pt\hbox{$_{_{\bullet}}$}}\over{#1}}
\def\un#1{\underline{#1}} \def\ul#1{\underline{#1}}

\def\plpl{{+\!\!\!\!\!{\hskip 0.009in}{\raise -1.0pt\hbox{$_+$}}
{\hskip 0.0008in}}}
\def\mimi{{-\!\!\!\!\!{\hskip 0.009in}{\raise -1.0pt\hbox{$_-$}}
{\hskip 0.0008in}}}


\vspace{20pt}
\begin{center}

{\Large\bf Supersymmetric Yang-Mills Equations in 10+2 Dimensions }\\

\vspace{30pt}

{\large Hitoshi ~Nishino\footnote
{\baselineskip=11pt Research supported in part by NSF Grant 
PHY-9341926 and DOE Grant DE-FG02-94ER4085}} \\
\vspace{7pt}
{\baselineskip=12pt
\it Department of Physics \\
University of Maryland \\
College Park, MD 20742-4111, USA \\
E-Mail: nishino@umdhep.umd.edu \\ }

\vspace{10pt}
and \\
\vspace {10pt}

{\large Ergin ~Sezgin\footnote
{\baselineskip=11pt Permanent address: Center for Theoretical Physics,  
Texas A\&M University, College Station, 
TX 77843}$^,$\footnote{\baselineskip=11pt Research supported in part by 
NSF Grant PHY-9411543}} \\
\vspace{7pt}
{\baselineskip=12pt
\it International Center for Theoretical Physics \\
P.O.~Box 586, I-34014, Trieste, Italy \\
E-mail: sezgin@phys.tamu.edu \\}

\vspace{1.0truecm}

\vspace{25pt}

\centerline{ABSTRACT}
\end{center}

{\baselineskip=15pt 

We present a model for supersymmetric Yang-Mills theory
in 10+2 dimensions.  Our construction uses a constant null vector,
and leads to a consistent set of field equations and constraints.
The model is invariant under generalized translations and
an extra gauge transformation. Ordinary dimensional reduction
to ten dimensions yields the usual supersymmetric Yang-Mills equations,
while dimensional reduction to 2+2 yields supersymmetric Yang-Mills
equations in which the Poincar\'e supersymmetry is reduced
by a null vector.  We also give the corresponding formulation 
in superspace.  

} 

\vfill\eject 

\hoffset=-.4in
\voffset=-.8in

\baselineskip=18.5pt

\setcounter{page}{2}

\section{Introduction}

      It has been proposed by Vafa \cite{Vafa1} that the Type IIB and
the Type I or SO(32) heterotic strings which do not admit a direct M-theory
unification in 10+1 dimensions, may in fact arise from a unifying theory in
$10+2$
dimensions, called F-theory. This picture emerged in the analysis of
certain Type
IIB string vacua. In a somewhat related fashion, Hull \cite{Hull} has also
proposed a twelve dimensional (12D) picture, but in $11+1$ dimensions. 
In another development,
Tseytlin \cite{t} has suggested that the worldvolume vector fields of the
Type IIB
Dirichlet three-brane in 10D may provide two extra dimensions to imply a
three-brane in 12D with $(11,1)$ signature. This idea has been realized
\cite{jr},
at least in the bosonic sector, and without the full 12D Poincar\'e 
invariance.\footnote{\baselineskip=11pt Long ago Nahm \cite{Nahm} 
showed that supergravity theories are
impossible in more than $10+1$ dimensions, and supersymmetric Yang-Mills
theories
in more than $10+1$ dimensions, because higher spin fields set in. In
\cite{Castellani}, it was shown that although the bose and fermi degrees of
freedom
could be matched in $11+1$ dimensions, no corresponding supergravity model
existed. The lack of full 12D Poincar\'e invariance 
can avoid these problems.}
However, the relation between the conjectured 12D theories 
with $(11,1)$ versus $(10,2)$ signatures is not clear at present.

As far as a three-brane in $10+2$ dimensions is concerned, the possibility
of its
existence was conjectured some time ago by Blencowe and Duff \cite{Duff}, 
on the basis of matching bose  and fermi degrees of freedom. There is a
subtlety, however, in $10+2$ dimensions, namely the minimal 
chiral superalgebra with $32$ component Majorana-Weyl spinors takes the form
\be
\{ Q_\a, Q_\b\} = (\g^{\m\n})_{\a\b}\, P_{\m\n}
+ (\g^{\m_1...\m_6})_{\a\b}\, Z_{\m_1...\m_6}\ ,
\la{12dalg}
\ee
where the $66$ generators $P_{\m\n}$ and the $462$ generators
$Z_{\m_1...\m_6}$ have only nonvanishing commutators with the 
Lorentz generators $M_{\m\n}$, and the Dirac $\g$-matrices are 
chirally projected. (For a discussion
of how this algebra might play a role in the classification 
of perturbative and nonperturbative multiplets of string theory, 
see \cite{Bars}.)  Now this algebra
does not contain the usual translation generators. Therefore, one does not
expect  a Poincar\'e invariant field theory of the usual kind, and strictly
speaking the usual argument about $d-p-1$ translational zero modes does
not really hold. Nevertheless, there may indeed exist evidence for the
conjecture
of \cite{Duff}, in view of some recent developments to which we now turn.

In an attempt to give a concrete description of M-theory based on
\cite{Vafa3}, Kutasov and Martinec \cite{Martinec} have proposed an
$N=(2,1)$ superstring theory, where the left movers
with local $N=2$ worldsheet supersymmetry live in $2+2$ dimensional target
space,
and right movers with local $N=1$ worldsheet supersymmetry surprisingly live
in $10+2$ dimensional target space \cite{Vafa3}. Thus the geometrical
spacetime can
at most be $2+2$ dimensional. There is however a null vector in the
construction, which can be chosen in two different ways, corresponding to
$1+1$ or $2+1$ dimensional target space. The resulting $1+1$ and 
$2+1$ dimensional target space theories are then proposed to be the 
worldvolume theories for  a 10D superstring or 11D supermembrane. 
Just as there is an underlying $2+2$ worldvolume with two kinds of 
null reductions, the target space of the resulting extended
object is also conjectured to be $10+2$ dimensional with different 
kinds of null reductions.  In \cite{Martinec}, 
it is furthermore speculated that the relevant
worldvolume theory  may involve a $10+2$ dimensional Yang-Mills theory,
presumably  reduced to $2+2$ dimensions.  
Motivated by these developments, and as a first step towards the
investigation of supersymmetric field theories in $10+2$ dimensions, 
in this paper we present a model of supersymmetric Yang-Mills theory 
by utilizing a constant null vector.
The usage of a null vector $n_\m$ in our construction  results in a
modified form
of the algebra (\ref{12dalg}), namely
\be
\{ Q_\a, Q_\b\} = (\g^{\m\n})_{\a\b}\, n_\n \  P_\m\ .
\la{12dmalg}
\ee
Due to this constant null vector throughout the formalism, the model will
lack the full Poincar\'e invariance in $10+2$ dimensions. 
The model has a generalized  
translational and extra gauge invariances,  and it reduces to the 10D
supersymmetric Yang-Mills system upon ordinary dimensional reduction. If we
relax the assumption of ordinary dimensional reduction, 
the model curiously reduces, modulo global matters 
that may involve Wilson lines, to the  10D supersymmetric
Yang-Mills equations depending only  on one combination of the two  extra
coordinates, in addition to the 10D coordinates.

In what follows, we will first describe the model. We will then examine its
dimensional reduction to $9+1$ and $2+2$ dimensions, followed 
by the superspace formulation of the model, and our conclusions.

\section{Supersymmetric Yang-Mills Model in $10+2$ Dimensions}

The field content we consider is the same as 10D supersymmetric
Yang-Mills multiplet, namely a real vector $~A_\m{}^I$~ in the adjoint
representation, and a Majorana-Weyl fermion $~\l^I$~ with the positive
chirality, $~\g_{13} \l^I = + \l^I$~, also in the adjoint representation 
labelled by the
indices $~{\scst I,~J,~\cdots}$. We propose the following supersymmetry
transformation rules:
\bea
\d_Q A\du\m I &=& \bar\e\g_\m\l^I ~~,
\la{trsfa} \\
\d_Q  \l^I &=& \fracm 1 4 \g^{\m\n\r} \e F_{\m\n}{}^I n_\r ~~,
\la{trsfb}
\eea
where $~n^\m$~ is a constant vector. The Dirac $\g$-matrices obey the
$SO(10,2)$
Clifford algebra, and the signature of spacetime is taken to be $(- + +
\cdots + -)$. Note
that ~$\g^\m$~ flips the chirality, and that the supersymmetry parameter
and the gauge
fermions have opposite chiralities. The above  ans\"atze for the
transformation rules are
motivated by the chirality properties of the fermions and the requirement
of translation
emerging in the commutation of  two supersymmetry transformations. These
requirements also
allow the possibility of taking the supersymmetry parameter and the gauge
fermion to have the same chirality, and the introduction of
the constant null vector in $\d_Q A_\m$. However,
we have realized that in this case the closure of the supersymmetry algebra
imposes too strong conditions on the fields to allow acceptable set of
equations. In particular, we obtain the constraint $n_{\[\rho}\,
F_{\m\n\]}=0$, which is
clearly too
strong.

Our guiding principle now is to establish the closure of the
supersymmetry transformation rules  (\ref{trsfa}) and (\ref{trsfb}).
In what follows, we shall first describe the full system of equations of
motion, constraints and symmetries, which together with (\ref{trsfa})
and (\ref{trsfb}) characterize fully our model.
Next, we shall explain the derivation of these equations.

To begin with, the constant vector $n^\mu$ must satisfy the condition
\be
n_\m n^\m = 0 ~~.
\la{null}
\ee
The Yang-Mills field obeys the field equation
\be
D_\m F^\m{}_{\[\r}{}^I n_{\s\]} + \fracm1 4 f^{I J K} \left( \bar\l^J
\g_{\r\s} \l^K \right) = 0  ~~,
\la{ymeq}
\ee
and the constraint
\be
n^\m F_{\m\n}{}^I = 0 ~~.
\la{nf}
\ee
The gauge fermion obeys the equation
\be
\g^\m D_\m \l^I \equiv \g^\m \left(\partial_\m \l^I + f^{I J K}
A\du\m J \l^K  \right)  = 0 ~~,
\la{lambdaeq}
\ee
and the constraints
\bea
& & n^\m D_\m \l^I = 0 ~~,
\la{ndlambda} \\
& & n^\m \g_\m \l^I =0 ~~.
\la{nglambda}
\eea
The whole system is invariant under the supersymmetry transformations
(\ref{trsfa}) and (\ref{trsfb}), the usual Yang-Mills
gauge transformations and the following extra local gauge transformation
\be
\d_\Omega A\du\m I = \Omega^I n_\m ~~,
\la{extratrsf}
\ee
subject to the condition
\be
n^\m D_\m \Omega^I(x) = 0 ~~.
\la{ndomega}
\ee

The commutator of two supersymmetry transformations yields a generalized
translation, the usual Yang-Mills gauge transformation and an extra gauge
transformation with parameters $\xi^\mu$, $\Lambda$, $\Omega$,
respectively, as follows:
\be
\[ \d_Q(\e_1), \d_Q(\e_2) \] = \d_\xi +\d_\Lambda + \d_\Omega\ ,
\la{closure}
\ee
where the composite parameters are given by
\bea
\xi^\m &=&  \bar\e_2 \g^{\m\n}\e_1\ n_\n\ , \la{xi}\\
\Lambda^I &=& -\xi^\m\ A_\m{}^I\ , \la{l}\\
\Omega^I &=& \half \bar\e_2 \g^{\m\n}\e_1\ F_{\m\n}{}^I\ .  \la{o}
\eea
Note that the global part of the algebra  (\ref{closure}) is given by
(\ref{12dmalg}).

We now explain the derivation of the above equations. First,
we note that the supersymmetry transformations close on the
gauge field as in (\ref{closure}).  Next, we find that
the commutator of two supersymmetry transformations on the
gauge fermion yields the result
\bea
\[ \d_Q (\e_1), \d_Q (\e_2) \] \l^I  &=&
\xi^\m D_\m \l^I - \fracm1{32} \left( 3\g_{\r\s}{}^\t
- 8 \g_\r \d_\s^\t \right)\ n_\t\z^{\r\s}\ \dsl \l^I  \nn \\
& &  -\fracm 7{16} \g_\r \nsl D_\s \l^I \z^{\r\s} + \fracm 3{32} \g_{\r\s}
n^\m D_\m \l^I \z^{\r\s} \nn \\
& & - \fracm1{32(6!)} \, \g^\n \g^{\r_1\cdots\r_6}\left( \nsl D_\n \l^I
- n_\n \dsl \l^I \right) \ \xi_{\r_1\cdots\r_6}  \ ,\la{lclosure}
\eea
where $~\xi^{\m\n}\equiv\left(\bar\e_2\g^{\m\n}\e_1\right)$,
$~\xi^{\m_1\cdots\m_6}\equiv\left(\bar\e_2\g^{\m_1\cdots\m_6}\e_1
\right)$, $~\nsl\equiv\g^\m n_\m$~ and  $~\dsl\equiv\g^\m D_\m$~.
In obtaining the above result, we have used the Fierz identity for four
Majorana-Weyl spinors of the same chirality in 12D:
\bea
&&\left(\bar\psi_1\psi_2 \right) \left(
\bar\psi_3\psi_4 \right) = -\fracm 1{32} \sum_i \left(\bar\psi_1 {\cal O}_i
\psi_4 \right) \left(\bar\psi_3 {\cal O}_i\psi_2\right) ~~, \nn\\
&&\{ {\cal O}_i \} \equiv \left\{ I, \fracm i{\sqrt{2!}} \g^{\m\n},
\fracm 1{\sqrt{4!}} \g^{\m_1\cdots\m_4} , \fracm i{\sqrt{6!}}
\g^{\m_1\cdots\m_6} \right\}~~, \la{Fierz}
\eea
and the following $\g$-matrix identities
\be
\g^{\m\n\r} \g_{\s\t} \g_\m = 6\g_{\s\t}{}^{\n\r}
+ 32\d\du{\[\s}{\[\n} \g\du{\t\]}{\r\]} - 20 \d\du{\[\s}\n \d\du{\t\]}\r~~,
\la{onegamma}
\ee
\be
\g^\m\g_{\n_1\cdots\n_6} \g_\m \equiv 0~~.
\la{sixgamma}
\ee

From (\ref{lclosure}) we see that the first term is the translation
$~\d_\xi$~ and gauge transformation $\d_\Lambda$ consistent with
(\ref{closure}), while the second, the third and the last term vanish
on-shell, when the $~\l\-$field satisfies the field equation
(\ref{lambdaeq}). The
two terms proportional to $\nsl$ vanish by imposing the constraint
(\ref{nglambda}), and the remaining term proportional to
$n^\m D_\m \lambda^I$ vanishes by imposing the constraint (\ref{ndlambda}).
It follows from (\ref{nglambda}) that the constant vector $n_\m$
must be null, {\it i.e.}, it must satisfy (\ref{null}).

We next check the invariance of the field equation (\ref{lambdaeq}) and the
constraints (\ref{ndlambda}) and (\ref{nglambda}) under supersymmetry
and extra gauge transformations. By varying (\ref{ndlambda}) we obtain
\be
\d_Q \big(n^\m D_\m \l^I \big) = \fracm 1 4 \g^{\r\s\t}
\e \left( - 2 D_\r F\du{\s\m} I \right) n^\m n_\t + f^{I J K} \l^K
\left( \bar\e\nsl \l^K  \right) ~~,
\ee
where we have also used the usual Bianchi identity $~D_{\[\m} F_{\n\r\]}{}^I
\equiv 0$. Now the second term vanishes under (\ref{nglambda}),
while for the first term  to vanish, we need the new constraint (\ref{nf}).
Similarly the variation of the constraint (\ref{nglambda}) yields
\be
\d_Q \big( n^\m \g_\m \l^I \big)
=  \half \g^{\n\r} \e F\du{\s\n}I n_\r n^\s + \fracm 1 4 \g^{\m\n}
\e F\du{\m\n}I n_\r n^\r  ~~,
\ee
which vanishes due to (\ref{nf}) and (\ref{null}).

The $~\d_Q\-$transformation of the constraint (\ref{nf}) is easily
confirmed:
\be
\d_Q\big( n^\m F_{\m\n}{}^I \big)
= \left(\bar\e \g_\n n^\m D_\m \l^I \right) - \left( \bar\e D_\n \nsl\l^I
\right)
= 0 ~~,
\ee
thanks to (\ref{ndlambda}) and (\ref{nglambda}). As for the Yang-Mills
field equation (\ref{ymeq}), it follows from the $~\d_Q\-$variation of the
$~\l^I\-$field equation
(\ref{lambdaeq}):
\be
0 = \d_Q \left( \g^\m D_\m \l^I  \right)
= \half \g^{\r\s} \e \left[ D^\n F_{\n\[ \r} {}^I n_{\s\]} +
\fracm 1 4 f^{I J K} \left( \bar\l^J \g_{\r\s} \l^K \right)  \right] ~~,
\ee
where we have used the constraint (\ref{nf}), together with
the Bianchi identity for $F_{\m\n}$. We have checked  that a further
supersymmetric variation of the Yang-Mills equation (\ref{ymeq}) vanishes
modulo the gauge fermion field equation (\ref{lambdaeq}),
and the constraints (\ref{nf}),
(\ref{ndlambda}) and (\ref{nglambda}).

At this point we have checked fully the supersymmetry of all the
equations of motion and the constraints. However, we still need to
check the invariance of
these equations under the extra gauge symmetry (\ref{extratrsf}).
Obviously, the
constraints (\ref{ndlambda})  and (\ref{nglambda}) are invariant under these
transformations, while the variation of the constraint (\ref{nf}) yields
\be
\d_\Omega \left( n^\n F\du{\n\m}I \right)  = \left(n^\n D_\n \Omega^I\right)
n_\m - \left( D_\m \Omega^I  \right) n^\n n_\n ~~.
\ee
The last term vanishes due to (\ref{null}), while the vanishing of the first
term requires the condition (\ref{ndomega}). The invariance of the
gaugino equation (\ref{lambdaeq}) is obvious due to the condition
(\ref{nglambda}).  Finally, we have also checked that
the Yang-Mills equation is invariant under the extra gauge transformation
(\ref{extratrsf}), thanks to the constraints (\ref{nf}) and (\ref{ndomega}).

To summarize, we have the supersymmetric Yang-Mills multiplet with the
supertransformation rule (\ref{trsfa}) and (\ref{trsfb}), and their field
equations (\ref{lambdaeq}) and (\ref{ymeq}). We need the null-vector 
$~n^\m$~ and other constraints on the fields given in
(\ref{nf}), (\ref{ndlambda}), (\ref{nglambda}) and (\ref{null}).
The on-shell closure of the gauge algebra
is guaranteed to produce the translation as well as the extra gauge
transformation (\ref{extratrsf}) with the condition (\ref{ndomega}) for its
parameter. The extra gauge transformation commutes with supersymmetry.

\section{Dimensional Reduction to Ten and Lower Dimensions}

In this section we shall use hats for the fields and indices in 12D, to be
distinguished from the unhatted ones in 10D. We choose our coordinates to be
$~(\Hat x^0, \Hat x^1, \cdots, \Hat x^9, \Hat x^{11},\Hat x^{12})$~ with the
metric $~(\Hat\eta_{\hat\m\hat\n} ) = \hbox{diag.}~(-+\cdots+-)$.
The 12D $\g$-matrices  satisfy $~\{\Hat\g_{\hat\m}, \Hat\g_{\hat\n}\} =
2 \Hat\eta_{\hat\m\hat\n}$. We can choose the $\g$-matrix to be
\be
\Hat\g_{\hat\m} = \cases{\hbox{$\Hat\g_\m = \g_\m \otimes \s_3 ~~,$} \cr
\hbox{$\Hat\g_{11} = I\otimes \s_1 ~~,$} \cr
\hbox{$\Hat\g_{12} = - I \otimes i\s_2 ~~.$} \cr }
\ee
Here $I$ is the $32\times 32$ unit matrix and the $\s$'s are
the Pauli matrices and $\g_\m$ are the $32\times 32$ $\g$-matrices
in $9+1$ dimensions. The charge conjugation matrix can be chosen as
\be
\Hat C= - \Hat\g_0 \Hat\g_{12}= \g_0 \otimes \s_1 \equiv C\otimes \s_1\ ,
\la{c}
\ee
where we have identified $\g_0$ with the charge conjugation matrix $C$ in
10D. Both $\Hat C$ and $C$ are antisymmetric. Next, we define the
matrix
\be
\Hat\g_{13} \equiv \Hat\g_0 \Hat\g_1 \cdots \Hat\g_9
\Hat\g_{11} \Hat\g_{12} = \g_{11} \otimes \s_3 ~~, \la{g13}
\ee
where $~\g_{11}\equiv \g_0 \g_1 \cdots \g_{9}$. Finally, we  choose the
null-vector as
\be
\left( \Hat n^{\hat\m} \right)
= (0, 0, \cdots, 0 , + 1, - 1) ~~, ~~~~ \left( \Hat n_{\hat\m} \right)
= (0, 0, \cdots, 0 , + 1, +1) ~~.
\la{nullvector}
\ee
Even though this form looks special, we can show that starting with an
arbitrary null-vector $~\Hat n'_{\hat\m}$, we can {\it always} utilize
orthogonal transformations within the two compact sub-manifolds to bring
its components to the standard form (\ref{nullvector}).

We are now ready to analyse the null reduction of our field equations and
constraints.  To begin with, the constraint (\ref{nglambda})
implies that $\s_+ \Hat \lambda=0$, where
$\s_\pm \equiv (\s_1\pm i\s_2) /{\sqrt 2}$.
Together with the 12D chirality condition, this implies that we can
write $\Hat\lambda $ as
\be
\Hat \lambda = \pmatrix{ \lambda \cr 0 \cr }\ , \qquad\qquad
\g_{11}\lambda=\lambda\ . \la{lambda}
\ee
Hence, the gauge fermion $\lambda$ is a Majorana-Weyl spinor of
$SO(9,1)$ as it should be in 10D supersymmetric Yang-Mills theory.
For simplicity in notation, we shall
use the matrix notation for the gauge field and gauge fermions, in the rest
of the paper.

Using (\ref{nullvector}) and (\ref{lambda}), we see that the remaining
constraints given in (\ref{nf}) and (\ref{ndlambda}) can be expressed as
follows:
\be
F_{-+}=0\ ,  \quad\quad
F_{-\m}=0\ , \quad\quad
D_-\lambda =0\ . \la{fl}
\ee
where we have used the coordinates $x^\pm\equiv(x^{11}\pm x^{12})/{\sqrt2}$.
Using (\ref{lambda}) and (\ref{fl}), we easily find that the field 
equations (\ref{ymeq}) and (\ref{lambdaeq}) take the form
\be
D^\m F_{\m\n}-  \bar \lambda \g_\n \lambda =0\ ,  \quad\quad
\g^\m D_\m \lambda=0\ .\la{ym10}
\ee
Furthermore, the supersymmetry transformation rules (\ref{trsfa}) and
(\ref{trsfb}) take the form
\be
\d_Q A_\pm = 0 ~~, \qq \d_Q A_\m = \bar\e\g_\m\l ~~,\qq
\d_Q  \l = \fracm 1 2 \g^{\m\n} \e F_{\m\n}  ~~. \la{tr}
\ee
Our model is also invariant under the following local Yang-Mills and
extra gauge transformations:
\bea
\delta A_+ &=& D_+ \Lambda +\Omega\ , \qq D_-\Omega =0\ ,  \la{gp}\\
\delta A_- &=& D_- \Lambda\ , \la{gm}\\
\delta A_\mu &=& D_\mu \Lambda\ , \qq
\delta \lambda = -\[\Lambda,\l \] \ .\la{gt}
\eea

It is interesting to note that we have obtained the 10D
supersymmetric Yang-Mills equations (\ref{ym10}) and supersymmetry
transformation rules (\ref{tr}), without any assumption on the dependence of
the fields on the extra two coordinates $x^\pm$.

Considering the so called ordinary dimensional reduction, we set
$\partial_+=\partial_-=0$.
In that case, using the local $\Omega$ transformations we can gauge away
$A_+$. Then, the first
equation in (\ref{fl}) is satisfied, while the second one reduces to $D_\mu
A_-=0$. From this
equation we deduce that $A_-$=0.  Then the last equation in (\ref{fl}) is
also satisfied, and the
full system of equations of motions and symmetry transformations reduces
precisely to the
10D supersymmetric Yang-Mills system.

If we do not assume independence of fields of the coordinates
$x^\pm$, we find that we can gauge away the
fields $A_\pm$, {\it modulo global issues}, by using the first constraint in
(\ref{fl}), together with the
$\Omega$ gauge transformation (\ref{gp}) and the $x^-$ dependent part of the
$\Lambda$ gauge transformation. Furthermore, using the second and third
constraints in (\ref{fl}), we observe that the fields $A_\m$
and $\l$ depend on the 10D spacetime coordinates and $x^+$ alone.

In summary, we obtain the 10D
supersymmetric Yang-Mills system with an arbitrary
dependence on the extra coordinate $x^+$.
From the 10D point of view, we can interpret  the coordinate
$x^+$ as the affinization of the Yang-Mills gauge group. We shall comment
further
on this point at the conclusions.

Let us now consider the ordinary dimensional reduction of the 12D model
down to $2+2$ dimensions, where we will see that the coordinates $x^+$ will
have a
different significance, since they will be treated as part for the $2+2$
dimensional spacetime.  We label the coordinates as $\Hat
x^{\hat\m}=(x^\mu, x^i)$
with
${\scst \m~=~0,~1,~11,~12}$ and ${\scst i~=~2,~...,~9}$, and
set $\partial_i=0$. We work with the  Dirac $\g$-matrices: $\Hat\g_\m = 
\g_\m \otimes I$ and $\Hat \g_i = \G_5\otimes \g_i$, where $\g_\m$ and 
$\g_i$ are the $SO(2,2)$ and $SO(8)$ Dirac matrices, respectively: $\G_5=
\g_0\g_1\g_{11}\g_{12}$, and $I$ is the $16\otimes 16$ unit matrix. It
follows that $\Hat\g_{13}=\G_5\otimes \G_9$ with $\G_9\equiv \g_2\cdots
\g_9$ and $\Hat C=C\otimes I$, where $C$ is the antisymmetric charge
conjugation matrix in $2+2$ dimensions.

The 12D chirality
condition on the gauge fermion becomes $\G_5 \G_9 \l =\l$. This means that
in $2+2$ dimensions we have a left-handed spinor in the ${\bf 8}_{\rm S}$ 
representation of $SO(8)$, and a right-handed spinor in the
${\bf 8}_{\rm C}$ representation. These make up $32$ real components. The
null-ness condition (\ref{nglambda}), and the Dirac equation further 
reduce the number of degrees of freedom to $8$, which is to be expected 
in an $N=4$ supersymmetric Yang-Mills multiplet in 4D. 
Note that the chirality condition
$\G_5 \G_9 \e =-\e $ implies that the left-handed parameters are in the 
${\bf 8}_{\rm C}$ representation of $SO(8)$, while the right-handed 
parameters are in the ${\bf 8}_{\rm S}$ representation.

Defining $A_i \equiv \phi^i$ and choosing the null-vector as
\be
        (n_\mu )=(0,0,1,1)\ , \quad\quad (n_i)=(0,...,0)\ , \la{nv1}
\ee
we find that the field equations of $10+2$ dimensions reduces to $2+2$
dimensions as follows:
\bea
&&D_\m F^\m{}_{\[\n}\, n_{\rho\]}-\[\phi_i,D_{\[\n} \phi^i\]\, n_{\rho\]}
+\fracm 1 2 \bar \l \g_{\n\rho}\l = 0\ ,
\quad\quad\quad\ n^\m F_{\m\n}=0 \ , \nn\\
&& D_\m D^\m \phi^i n_\n + \[\phi_k,\[\phi^k,\phi_i\]\]\, n_\n -
\bar\l \g_\n\g_i\G_9\l = 0
\ , \quad\quad n^\m D_\m \phi_i=0\ , \nn\\
&&\g^\m D_\m \l +\g^i \[\phi_i,\l\]=0\ ,\quad\quad n^\m D_\m \l=0\ ,
\quad\quad n^\m \g_\m \l=0\ .  \la{nle}
\eea
These equations are invariant under the following supersymmetry
transformations
\bea
\delta_Q A_\m &=& \bar\e \g_\m \l\ ,\quad\quad
\delta_Q \phi_i = \bar\e \g_i\G_9 \l\ , \nn\\
\delta_Q \l &=& \left( \fracm 1 4 \g^{\m\n}\
F_{\m\n}-\fracm 1 2 \g^\m\g^i\G_9
D_\m\phi_i +\fracm 1 2 \g^{i j} \phi_i\phi_j \right) \nsl \e \ . \la{nst}
\eea
In these equations we have suppressed the $SO(8)$ indices. A simple way to
make them
explicit is to split the fermions into their left and right-handed
$SO(2,2)$ chiralities,
and then to label the left (right)-handed gauge fermions with ${\scst A
~(\Dot A)}$, and the left (right)-handed supersymmetry parameter
with ${\scst\Dot A ~(A)}$, where
${\scst A,~\Dot A ~=~1,~...,~8}$. For example,
$\delta \phi^i = \bar\e^A (\g^i)_{A\Dot A}\l^{\Dot A}
- \bar\e^{\Dot A} (\g^i)_{\Dot A A}\l^A$, with the $SO(2,2)$ chiralities
suppressed,
but clearly correlated with the $SO(8)$ chiralities.

The dimensionally reduced model also has an extra gauge symmetry under
which only the
$A_\mu$ transforms:  $\delta A_\m= \Omega n_\m $. The model is
a version of $N=4$ super Yang-Mills that has an $SO(1,1)\otimes
SO(8)$ bosonic symmetry, living in $2+2$ dimensional spacetime.

We can perform a different kind of dimensional reduction to $2+2$
dimensions, in which
again $\partial_i=0$, but with the null vector chosen as
\be
(n_\m)=(0,0,0,1)\ ,\qquad\qquad (n_i)=(1,0,...,0) \ . \la{nv2}
\ee
This will lead to a version of $N=4$ super Yang-Mills which has
$SO(2,1)\otimes SO(7)$
bosonic symmetry, but lives in $2+2$ dimensions. It is straightforward to
obtain the analogs of (\ref{nle}) and (\ref{nst}) for this case, but we
shall skip the
details here.

\section{Superspace Formulation}

We have thus far established the component formulation for 12D 
supersymmetric Yang-Mills theory.  Our next natural task is to 
re-formulate the same system
in superspace.  We have a chiral superspace with coordinates $Z^M$, and
supervielbein $E_M{}^A$, where the tangent space indices are 
~${\scst A ~=~(a,\a)}$~ with
~${\scst a ~=~0,~1,~2,~\cdots,~9,~11,~12}$~ labelling the bosonic directions,
and ~${\scst\a~=~1,~\cdots, ~32}$~ labelling
the fermionic directions.  We may of course have spinor superfields that
carry spinor indices of opposite chirality labelled by
~${\scst \Dot\a~=~1,~...,~32}$~.  We use the (anti)
symmetrization symbols such as $~{\scst\[ A B )}$~ with unit strength
normalization in superspace. Our superspace conventions are those of
\cite{bst}.

We proceed by defining the torsion super two-form $T^A=dE^A$, as can be
read from the
superalgebra (\ref{12dmalg}):
\be
T^c = e^\a \wedge e^\b\, (\s^{cd})_{\a\b}\, n_d\ , \quad\quad  T^\a =0\ ,
\la{tc}
\ee
where the basis one-forms are defined as $e^A=d Z^M E_M{}^A$.
Hence, they satisfy
\be
d e^c= e^\a \wedge e^\b\ (\s^{cd})_{\a\b}\ n_d\ ,\quad\quad d
e^\a = 0\ . \la{se}
\ee

We next define the Yang-Mills curvature super two-form $F=d A+A^2$ 
as follows
\be
F=e^\a \wedge e^b \left[\,  n_b \chi_\a - 2(\s_b\l)_\a\, \right]
+ \fracm1 2 e^a \wedge e^b\
F_{b a}\ , \la{fc}
\ee
where we have introduced the chiral spinor superfield $\chi_\a$ and the
anti-chiral spinor superfield $\l_{\Dot\a}$. The presence of the two-form 
superfield $F_{a b}$ is as
expected, and its zero-th order in $\theta$ component is the usual
Yang-Mills field
strength. The presence of the spinor superfield
$\l_{\Dot\a}$ is also as expected, and its zero-th order in $\theta$
component is the
gauge fermion. However, the occurrence of the spinor superfield $\chi$ is
special to our
model. It is needed for the closure of the super two-form $F$, but it drops
out of the
physical field equations, as we will explain further below.

Our task is to show that $D F =0$, modulo the field equations 
and constraints of the
previous section. To this end, we also impose the following constraints on
various quantities occurring in (\ref{fc}):
\bea
&& n^a n_a=0\ , \quad\quad  n^a F_{a b} = 0 \ , \la{s1}\\
&& n^a (\s_a)_\a{}^{\Dot\b} \l_{\Dot\b} = 0 \ , \quad\quad n^a\nabla_a
\l_{\Dot\b}= 0 \ , \la{s2}\\
&& \nabla_\a \l_{\Dot\b} = -\fracm1 4 (\s^{c d e})_{\a\Dot\b}
F_{c d}\ n_e  \ , \quad\quad
\nabla_{(\a}\chi_{\b)} = \fracm1 2( \s^{a b} )_{\a\b} F_{a b} ~~, \la{s3}\\
&& \nabla_\g F_{a b}  = 4\left( \s_{\[a} \nabla_{b\]} \l\right)_\g -
                          2 n_{\[a} \nabla_{b\]}\chi_\g ~~.  \la{s4}
\eea

We now show that $D F=0$. First, we observe that the terms in $D F $
proportional to
$e^\a \wedge e^\b \wedge e^\g$ vanish due to the identity
\be
(\s_a{}^b)_{(\a\b} (\s^a)_{\g)}{}^{\Dot\d}=
\fracm1{12} (\s_{c d})_{(\a\b} (\s^{c d}\ \s^b)_{\g)}{}^{\Dot\d} ~~,\la{id1}
\ee
together with the first of the constraints in (\ref{s2}). The superfield
$\chi_\a$ drops
out automatically in this sector. We next examine the terms in $D F $
proportional to
$e^\a \wedge e^\b \wedge e^c$. We find that these terms vanish due to the
constraints
(\ref{s3}). It is in this sector that we find the need to introduce the
superfield
$\chi_\a$. Next, we find that the terms in $D F $ proportional to
$e^a \wedge e^b \wedge e^\g$
vanish, due to the constraint (\ref{s4}). Finally, the terms proportional to
$e^a \wedge e^b \wedge e^c$ vanish trivially,
due to the usual Bianchi identity $D_{\[a}F_{bc\]}=0$.

The constraints (\ref{s1})-(\ref{s4}) encode the information about the
supersymmetric
Yang-Mills field equations. We can use (\ref{s4})  to get the $~\l\-$field
equation by
evaluating the anti-commutator $~\{\nabla_\a,\nabla_\b \} \l_{\Dot\g} $~ in
two  different
ways:
\be
T\du{\a\b}c \nabla_c \l_{\Dot\g}  = \left\{\nabla_\a,
\nabla_\b\right\} \l_{\Dot\g}
= 2 \nabla_{(\a}\left(\nabla_{\b)} \l_{\Dot\g} \right) ~~,\la{leq1}
\ee
from which we find the gauge fermion field equation
\be
(\s^a)_\b{}^{\Dot\g} \nabla_a\l_{\Dot\g} = 0 ~~,  \la{le3}
\ee
in agreement with \eqref{lambdaeq}.
In this process we also see that the superfield $~\chi$~
disappears completely, and it plays only a role of auxiliary superfield.

By taking a spinorial derivative of (\ref{le3}), we can get the
Yang-Mills  field equation
\be
\nabla_a F^a{}_{\[ b}\  n_{c \]}+\fracm1 2(\s_{b c})_{\Dot\a}{}^{\Dot\b}\,
                                 \l^{\Dot\a}\l_{\Dot\b}=0 ~~.   \la{leq2}
\ee
Interestingly, the superfield
$~\chi$~ disappeared completely also from this equation. We also note that
in deriving \eqref{leq2}, we have used the following identity:
\be
(\s^a)_{(\a}{}^{\Dot\g} (\s^a)_{\b)}{}^{\Dot \d} = \fracm1 8
(\s^{b c})_{\a\b} (\s_{b c})^{\Dot\g\Dot\d} ~~. \la{id2}
\ee

Finally, we mention the consistency related to $~\chi\-$superfield.  We can
confirm the consistency of $~\{\nabla_{(\a} , \nabla_\b \}\chi_{\g)}
= 2 \nabla_{(\a}\left( \nabla_{\b} \chi_{\g)} \right)$ by comparing
the two sides, and see they actually coincide after the use of the
$~n\-$dependent constraints, as well as the identity \eqref{id1}.

\section{Conclusions}

In this paper we have given an explicit field theoretic realization
of the 12D superalgebra (\ref{12dmalg}), in terms of
supersymmetric Yang-Mills multiplet. The key
to our construction is the use of a constant null vector, as motivated by
$(2,1)$ strings \cite{Vafa3,Martinec}. The resulting model,
of course, lacks the full Poincar\'e invariance in $10+2$
dimensions, as expected in F-theory
\cite{Vafa1}, and as is evident from the underlying algebra
(\ref{12dmalg}).
                                        
We have shown that the ordinary dimensional reduction to 10D 
yields the usual supersymmetric Yang-Mills equations, 
while two kinds of ordinary dimensional reductions 
to $2+2$ dimensions yield new kinds of $N=4$ super Yang-Mills systems
where the $SO(2,2)$ group is broken down to $SO(1,1)$ or $SO(2,1)$,
depending on the choice of the null vector, 
in accordance with \cite{Vafa3,Martinec}.

We have also found the curious result that if we relax the restriction of
ordinary dimensional reduction, we obtain, modulo global
matters which may involve Wilson lines, the 10D supersymmetric
Yang-Mills system depending on the extra coordinate $x^+$. This dependence
can be interpreted as a straightforward affinization of the 
Yang-Mills group, which, however, appears to be redundant. 
The resolution of this issue may lie in finding a                           
suitable action principle for our model, which is lacking at present. An
appropriate superstring or super three-brane action may also                 
shed light on this issue, by providing the extra information about the 
dependence of the target space fields on the $x_\pm$ coordinates.

We have also established the corresponding superspace formulation, 
in which we see the necessity of the fermionic superfield $\chi$ that  
eventually disappears in the superfield equations for physical fields.  

A future direction to extend the present work is the construction of
supergravity theories in $10+2$ dimensions. A hint to the structure 
of these theories may come from the requirement of 
$\kappa$-symmetry of a possible Green-Schwarz type 
action in curved superspace. One approach would be to lift 
the 10D heterotic string action to $10+2$ dimension by utilizing a null
vector. However, a more satisfactory approach might be the construction of a  
three-brane action in which a $2+2$ dimensional worldvolume is embedded in    
$10+2$ dimensional spacetime. In either case, 
a curved superspace version of the flat superspace with 
a null vector described here may be relevant. We are
currently investigating these issues.

In conclusion, we stress that the 12D model presented here provides  a
unified description for a variety of interesting supersymmetric field
theories that
can be obtained by dimensional reduction. As far as F-theory is concerned,
although its low energy limit is not known with certainty at present, 
we hope that our results will be of relevance to this intriguing problem.

\vspace{8mm}
\noindent{\bf Acknowledgements}
\vspace{4mm}

We are grateful to M.J.~Duff, S.J.~Gates, Jr., T.~H\"ubsch, J.~Strathdee,
C.~Vafa and E.~Witten for helpful discussions.  E.S.~would like to thank
the International Center for Theoretical Physics 
in Trieste for hospitality.

\pagebreak

\end{document}